\documentstyle[prl,aps,epsfig,multicol]{revtex}
\newcommand{\bm}{\bibitem}

\def\be {\begin{equation}}
\def\ee {\end{equation}}
\def\bea {\begin{eqnarray}}
\def\eea {\end{eqnarray}}
\def\nn {\nonumber}

\begin{document}
\title{Energy loss and dynamical evolution of quark $p_T$ spectra }
\author{Pradip Roy$^a$, Abhee K. Dutt-Mazumder$^a$ and Jan-e Alam$^b$ } 
\address{a) Saha Institute of Nuclear Physics, 1/AF Bidhannagar, Kolkata, India}
\address{b) Variable Energy Cyclotron Centre, 1/AF Bidhannagar, Kolkata, India}
\maketitle

\vspace{0.5cm}

\begin{abstract}
Average energy loss of light quarks has been calculated in a two stage
equilibrium scenario where the quarks are executing Brownian motion
in a gluonic heat bath.  The evolution of the quark $p_T$ spectra is
studied by solving Fokker-Planck equation in an expanding plasma. Results
are finally compared with experimentally measured pion $p_T$ spectrum
at RHIC.
\\[0.1 cm]
\end{abstract}
\vspace{0.2cm}
\begin{multicols}{2}

Production of high transverse momentum ($p_T$) particles in heavy ion 
collision, in recent years, has assumed special interest. This is related 
with the phenomenon of high $p_T$ particle suppression -- dubbed as
jet quenching. Actually energetic partons while passing through 
plasma lose energy which 
degrades the population of high $p_T$ hadrons. 
Experimentally such suppression has been observed 
at Relativistic Heavy Ion Collider (RHIC) at high $p_T$ domain~\cite{npa757}. 
This phenomenon of `jet quenching' can
be used to extract the properties of early stage of the plasma, 
temporarily produced in the high energy heavy ion collision~\cite{bjorken}.

To study the modified $p_T$ spectrum of hadrons it is essential
to estimate parton
energy loss in the thermal bath of quarks and gluons.  
Partons in a plasma can dissipate energy in two ways, 
either by two body scattering (collisional loss) or 
via the emission of gluons (radiative loss). 
Significant progress 
has been made in recent years to calculate partonic energy loss 
\cite{bjorken,baierjhep,baierreview,zakharov,gyulassy00,stephane,thoma91,abhee05}. 
In many of these calculations, first, path length dependent energy loss  
is estimated, which, consequently is used to modify the fragmentation 
functions that depopulate high $p_T$ hadrons~\cite{wangnpa}.
Present approach is different in the sense that here we dynamically evolve 
the quark $p_T$ spectra for a given initial distribution.   
Importance of this has 
already been discussed in refs.~\cite{baierjhep,jeon} where the
focus has been on the radiative loss. Even though this is the main mechanism of 
energy loss, under certain kinematic conditions,  
collisional loss could be comparable, or even be more, to its
radiative counterpart \cite{abhee05}. This is particularly so in case of heavy
quark because of the dead cone effect \cite{kharzeev}.

To estimate  
energy loss, we inject quarks with a given energy distribution  
and study the broadening of the same as the system expands and cools. 
In the present model gluons are thermalized at a much smaller time scale than
quarks. Such two stage equilibration scenario, albeit in a 
different context, was considered sometime ago \cite{shuryakprl92,alamprl94}.
Possibility of earlier thermalization (even faster)
of (soft) gluons is further accentuated, if 
the gluons are assumed to be in color glass condensate (CGC) state initially.
In CGC scenario, it has been shown that thermal bath for the gluons could 
be formed much earlier. Typical formation time scale for RHIC 
and LHC energies are estimated to be 
$t_i=1.40$ GeV$^{-1}$, $t_i=0.62$ GeV$^{-1}$ respectively \cite{rajuprc01}. 

Under such simplifying scenario, it thus reduces to the problem
where quarks are executing Brownian motion
in the gluonic heat bath. The evolution of the quarks are, therefore,
governed by the Fokker-Planck equation (FPE).  So we avoid solving full 
Boltzmann kinetic equation (BKE) and approximate relevant collision integral 
in terms of appropriately defined drag and diffusion coefficients
\cite{svetitsky,moore05,mustafa}. 

The drag and diffusion co-efficients are calculated using techniques of finite 
temperature perturbative quantum chromodynamics.  The former is related to 
the quark energy loss, while the latter is related to the square of the 
momentum transfer \cite{svetitsky,moore05}. Both of these quantities show 
infrared divergences as the 
collisions are dominated by soft ($t$-channel) gluons. 
To cure this problem, we screen the interaction via 
hard thermal loop (HTL) corrected propagator which makes long 
range Coulomb interaction finite. 
 
To arrive at the relevant FP equation from BKE we assume that there
is no external force and therefore,

\bea
\left ( \frac{\partial }{\partial t} +
{\bf v_p\cdot \nabla_r} \right )
f({\bf p,x},t) =C[f({\bf p,x},t)]
\label{eq:boltz}
\eea
Here, quarks have a phase space distribution which evolves in
time and the collision term is evaluated by considering ultra-relativistic
scattering of the quarks and gluons which eventually are expressed in terms
of transport coefficients.
Considering that the system expands in the longitudinal direction  
Eq.(\ref{eq:boltz}) takes the following form \cite{gordonplb}:

\bea
\frac{\partial f({\bf p},z,t)}{\partial t} +
v_{pz}\frac{\partial f({\bf p},z,t)}{\partial z} =C[f({\bf p},z,t)].
\eea
Here, $v_{pz}=p_z/E_p$ (for light partons $E_p=|p|$). This equation can
be simplified further for the central rapidity region which is boost
invariant in rapidity, which implies
\bea
f({\bf p_T},p_z,z,t)=f({\bf p_T},p_z^\prime,\tau).
\eea
Here $p_z^\prime=\gamma(p_z- u_z p)$, the transformation velocity
$u_z=z/t$, $\gamma=(1-u_z^2)^{-1/2}=t/\tau$ and $\tau=\sqrt{t^2-z^2}$ 
denotes the proper time. Using the Lorentz transformation relation
$\partial\tau/\partial z|_{z=0}=0$, $\gamma_{z=0}=1$  and 
$\partial p_z^\prime/\partial z|_{z=0}= -p/t$, one finds
\bea
v_{pz}\frac{\partial f}{\partial z}=-\frac{p_z}{t} 
\frac{\partial f}{\partial p_z}
\eea
Therefore the Boltzmann equation takes the following form
\bea
\frac{\partial f({\bf p_T},p_z,t)}{\partial t} |_{p_zt} =
\left (\frac{\partial}{\partial t}
-\frac{p_z}{t}\frac{\partial}{\partial p_z}\right )f({\bf p_T},p_z,t)
\eea
\bea
\left (\frac{\partial}{\partial t}
-\frac{p_z}{t}\frac{\partial}{\partial p_z}\right )f({\bf p_T},p_z,t)
=C[f({\bf p_T},p_z,t)].
\label{eq:boltzBj}
\eea

Evidently in Eq.~\ref{eq:boltzBj}, the second term on the left hand side
represents the expansion while the
right hand side characterizes the collisions. The latter  
can be written in terms of the differential collision rate $W_{{\bf p,q}}$
\bea
C[f({\bf p_T},p_z,t)]=\int d^3q [ 
W_{p+q; q} f({\bf p+q})
-W_{p; q} f({\bf p})
]
\eea
which quantifies the rate of change of the quark momentum from ${\bf p}$ to
${\bf p-q}$, $W_{{\bf p,q}}= d\Gamma({\bf p,q})/d^3q$, where 
$\Gamma$ represent scattering rates.

In a partonic plasma, small angle collisions, with parametric dependence of 
$O(g^2T)$, are more frequent than the large angle scattering rate. The
latter goes as $\sim O(g^4T)$.  Therefore the distribution function does not 
change much over the mean time between two soft scatterings. This allows
us to approximate $f({\bf p +q}) \simeq f({\bf p})$. In contrast, 
$W_{{\bf p,q}}$, being
sensitive to small momentum transfer, falls off very fast with increasing 
$q$. Therefore, we write

\bea
W_{{\bf p+q,q}}f({\bf p + q})
\simeq
W_{{\bf p,q}}f({\bf p})
+ q_i \frac{\partial}{\partial p_i} (W_{\bf p,q}f)\nn\\
+\frac{1}{2} q_i q_j  \frac{\partial^2}{\partial p_i \partial p_j} 
(W_{\bf p,q}f)
\eea

With these approximation, Eq.~\ref{eq:boltzBj} can be written as

\bea
\left (\frac{\partial}{\partial t}
-\frac{p_z}{t}\frac{\partial}{\partial p_z}\right )f({\bf p_T},p_z,t)
&&=\frac{\partial}{\partial p_i}A_i({\bf p}) f({\bf p}) +\nn\\ 
&&\frac{1}{2}
\frac{\partial}{\partial p_i \partial p_j}[B_{ij}({\bf p})f({\bf p})], 
\label{fpexp}
\eea
where we have defined the following kernels,
\bea
A_i=\int d^3q W_{\bf p,q} q_i
\eea
\bea
B_{ij}=\int d^3q W_{\bf p,q} q_i q_j
\eea

\bea
A_i&=&\frac{\nu}{16p(2\pi)^5}\int 
\frac{d^3{k^\prime}}{{k^\prime}}
\frac{d^3k}{k}
\frac{d^3{q}}{{p^\prime}}d\omega q_i
|{\cal{M}}|_{t\rightarrow 0}^2 f(k) (1+f(k{^\prime}))\nn\\ 
&&\delta^3({\bf q}-{\bf k^\prime}+ {\bf k})
\delta (\omega-{\bf v_{k^\prime}\cdot q})
\delta (\omega-{\bf v_k\cdot q})
\label{eq:drag1}
\eea

\bea
B_{ij}&=&\frac{\nu}{16p(2\pi)^5}\int 
\frac{d^3{k^\prime}}{{k^\prime}}
\frac{d^3k}{k}
\frac{d^3{q}}{{p^\prime}}d\omega q_iq_j
|{\cal{M}}|_{t\rightarrow 0}^2 f(k)\nn\\&&(1+f(k{^\prime})) 
\delta^3({\bf q}-{\bf k^\prime}+ {\bf k})
\delta (\omega-{\bf v_{k^\prime}\cdot q})
\delta (\omega-{\bf v_k\cdot q}),
\label{eq:diffusion}
\eea

First we consider Compton scattering ($gq(\bar{q})\rightarrow gq(\bar{q})$)
for which 
\bea
|{\cal M}|^2= g^4
\left [ 
\frac{s^2+u^2}{t^2}
-\frac{4}{9} \frac{s^2+u^2}{us}
\right] 
\label{eq:matsq}
\eea
In terms of explicit momentum variables it reduces to,
\bea
 |{\cal{M}}|_{t\rightarrow 0}^2 =8 g^4 
\frac{p^\prime k^\prime p k}{(q^2-\omega^2)^2}
(1-cos\theta_{pk}) (1-cos\theta_{p^\prime k^\prime }).
\eea
We label the incoming four momenta of the test and bath particles as
$P$ and $K$ respectively  
and the corresponding final momenta are $P^\prime$ and
$K^\prime$. In Eq.(\ref{eq:matsq}) $s,u$ and $t$ are Mandelstam variables with  
$t=\omega^2
-q^2$ and $s=-(u+t)$. The scattering angles, following ref.~\cite{moore00}, 
can be written as 
\bea
cos\theta_{pq}&=&\frac{\omega}{q}+\frac{t}{2pq}\nn\\
cos\theta_{kq}&=&\frac{\omega}{q}-\frac{t}{2pq}\nn\\
\eea

The parton parton collision is dominated by the soft momentum transfer.  
Therefore, we consider small angle scatterings to estimate the leading 
contribution, 
In this limit $t\rightarrow 0$ and 
$s=-u$. Hence the collision is dominated by the $t$ channel, 
$|{\cal M}|^2 \rightarrow g^4 2 s^2/t^2 $. We also have 
$cos\theta_{kq} \sim cos\theta_{k^\prime q} 
\sim \omega/q  
\sim cos\theta_{pq} \sim cos\theta_{p^\prime q}$ 
and
$cos\theta_{pk}=\omega^2/q^2 + (1-\omega^2/q^2)cos\phi_{pk}$. 
With these azimuthal angle averaged matrix element becomes,
\bea
\langle |{\cal M}| \rangle_{t\rightarrow 0}^2  
&&\simeq
12 g^4 p^\prime k^\prime p k
(1-\omega^2/q^2)^2/(q^2-\omega^2)^2\nn\\
&&\simeq
12 g^4 p^2 k^2 /q^4
\label{eq:avmatel}
\eea

It might be noted that 
Eq.~\ref{eq:drag1} is symmetric under $k\leftrightarrow k^\prime$, which 
allows one to write, 

\bea
A(p^2)&=&\frac{\nu}{16p^2(2\pi)^5}\int 
\frac{d^3{k^\prime}}{{k^\prime}}
\frac{d^3k}{k}
\frac{d^3{q}}{{p^\prime}}d\omega p\cdot q 
\langle {\cal M} \rangle_{t\rightarrow 0}^2\nn\\  
&&\frac{1}{2}[f(k) (1+f(k{^\prime}))- 
f(k^\prime) (1+f(k))]\nn\\ 
&&\delta^3({\bf q}-{\bf k^\prime}+ {\bf k})
\delta (\omega-{\bf v_{k^\prime}\cdot q})
\delta (\omega-{\bf v_k\cdot q}).
\label{eq:drag2}
\eea
recognizing the fact that $\omega$ is small compared to the momentum,
we have 
\bea
f(k)-f(k^\prime)=-\omega \frac{\partial f}{\partial k}.
\eea
 Furthermore,
\bea
&&-\int k^2 \frac{\partial f(k)}{\partial k} dk\nn\\
&&=\int 2 k f(k) dk =\frac{\pi^2 T^2}{3}.
\eea
With the help of these identities, the drag coefficient can easily be 
calculated at the leading log order: 
\bea
A{(p^2)}&=& \frac{\nu\pi\alpha_s^2 T^2}{6p}{\cal L},
\eea
 where ${\cal L}=
\int \frac{dq}{q}$ \cite{rajuprc01}. 
Evidently, ${\cal{L}}$ is infrared singular.  Such divergences do not arise 
if close and distant collisions are treated separately. 
For very low momentum transfer the concept of individual 
collision breaks down and one has to take collective excitations of
the plasma into account. 
Hence there should be a lower momentum cut off above which bare interactions
might be considered. While for soft collisions medium modified hard 
thermal loop corrected propagator should be used~\cite{thoma91,abhee05}. 
It is evident 
that Eq.~\ref{eq:drag1} actually gives $dE/dt$ or the energy loss rate
\cite{abhee05} that can be related to the drag coefficient.

$B_{ij}$ can be decomposed into  longitudinal
and transverse components: 
\bea
B_{ij}=B_t (\delta_{ij}-\frac{p_i p_j}{p^2}) + B_l \frac{p_ip_j}{p^2}
\eea
Explicit calculation shows that the off diagonal components of $B_{ij}$
vanish. 

\bea
B_{t,l}&&=\frac{\nu g^4}{(2\pi)^4}
\int\frac{d^3kd^3qd\omega}{2k2k^\prime 2p 2p^\prime}
\delta(\omega-{\bf v_p\cdot q})
\delta(\omega-{\bf v_k\cdot q})\nn\\
&&\langle {\cal M} \rangle_{t\rightarrow 0}^2
f(k)[1+f(k)+\omega\frac{\partial f}{\partial k}]
q_{t,l}^2.
\eea
Here, in the small angle limit $q_l\simeq \omega$ and 
$q_t \simeq \sqrt{q^2-q_l^2}$. 
With all these, in the leading log approximation 
\bea
B_t=\frac{2\nu \pi\alpha_s^2}{3}T^3{\cal L}
\eea
\bea
B_l=\frac{\nu \pi\alpha_s^2}{3}T^3{\cal L}
\eea
Here the fluctuation-dissipation theorem, is found to be satisfied 
automatically. This connects drag and momentum diffusion constants giving 
rise to Einstein's relation $B(p^2)= 2T E A(p^2)$, where 
$B_{ij}(p^2)=\delta_{ij}B(p^2)$.  To remove arbitrariness related to the 
momentum cutoff scales hidden in ${\cal L}$, hard thermal loop corrected 
propagator is used~\cite{abhee05}.

In the coulomb gauge, we can define $D_{00}=\Delta_l$ and
$D_{ij}=(\delta_{ij}-q^i q^j/q^2)\Delta_t$ where
$\Delta_l$ and $\Delta_t$ denote the longitudinal and transverse
gluon propagators given by,
\bea
\Delta_l(q_0,q)^{-1}=q^2-\frac{3}{2}\omega_p^2
\left [
\frac{q_0}{q}ln\frac{q_0+q}{q_0-q}-2
\right ]
\eea

\bea
\Delta_t(q_0,q)^{-1}=q_0^2-q^2+\frac{3}{2}\omega_p^2
\left [\frac{q_0(q_0^2-q^2)}{2q^3}
ln\frac{q_0+q}{q_0-q}-\frac{q_0^2}{q^2}
\right ]
\eea

With this, matrix element in the limit of small angle scattering,
we get the following
expression for the squared matrix element,
                                                                                  
\bea
|{\cal{M}}|^2&=& g^4 C_{qq} 16 (E E_1)^2
\vert \Delta_l(q_0,q)  \nn\\
&+& (v\times {\hat{q}}).(v_1 \times \hat{q})
\Delta_t(q_0,q)\vert^2
\eea
with $v=\hat{p}$, $v_1=\hat{p_1}$ and $C_{qq}$ is the color factor. 
With the screened interaction,  
the drag and diffusion constants can be calculated along the line of 
ref.~\cite{abhee05}. 

Now to calculate average energy loss of the light quarks
in an expanding partonic plasma we inject test quarks having following 
distribution at time $t_i$, 
\bea
f(p_T,p_z,t=t_i)=N\delta^2(p_T-p_{T0})\delta(p_z-p_{z0})
\eea
Bjorken cooling law~\cite{jdb83}, $\tau_i T_i^{1/{c_s^2}}=\tau T^{1/{c_s^2}}$ 
is used to describe the space-time evolution of the 
plasma. Here $c_s$ is the velocity
of sound.  

As the time progresses the system expands and $f(p_T,p_z,t)$ evolves
according to Eq.~\ref{fpexp}, which is solved numerically. 
The average energy loss is given by 
\bea
\langle \Delta E\rangle=\langle E \rangle-E_0
\eea
where $\langle E(t)\rangle=\int d^3p E f(p_T,p_z,t)/ \int d^3p  f(p_T,p_z,t)$. 
 $E_0$ corresponds to the peak of the initial spectrum, {\it i.e.}
$E_0=<E(t_i)>$.

\begin{figure}
\epsfig{figure=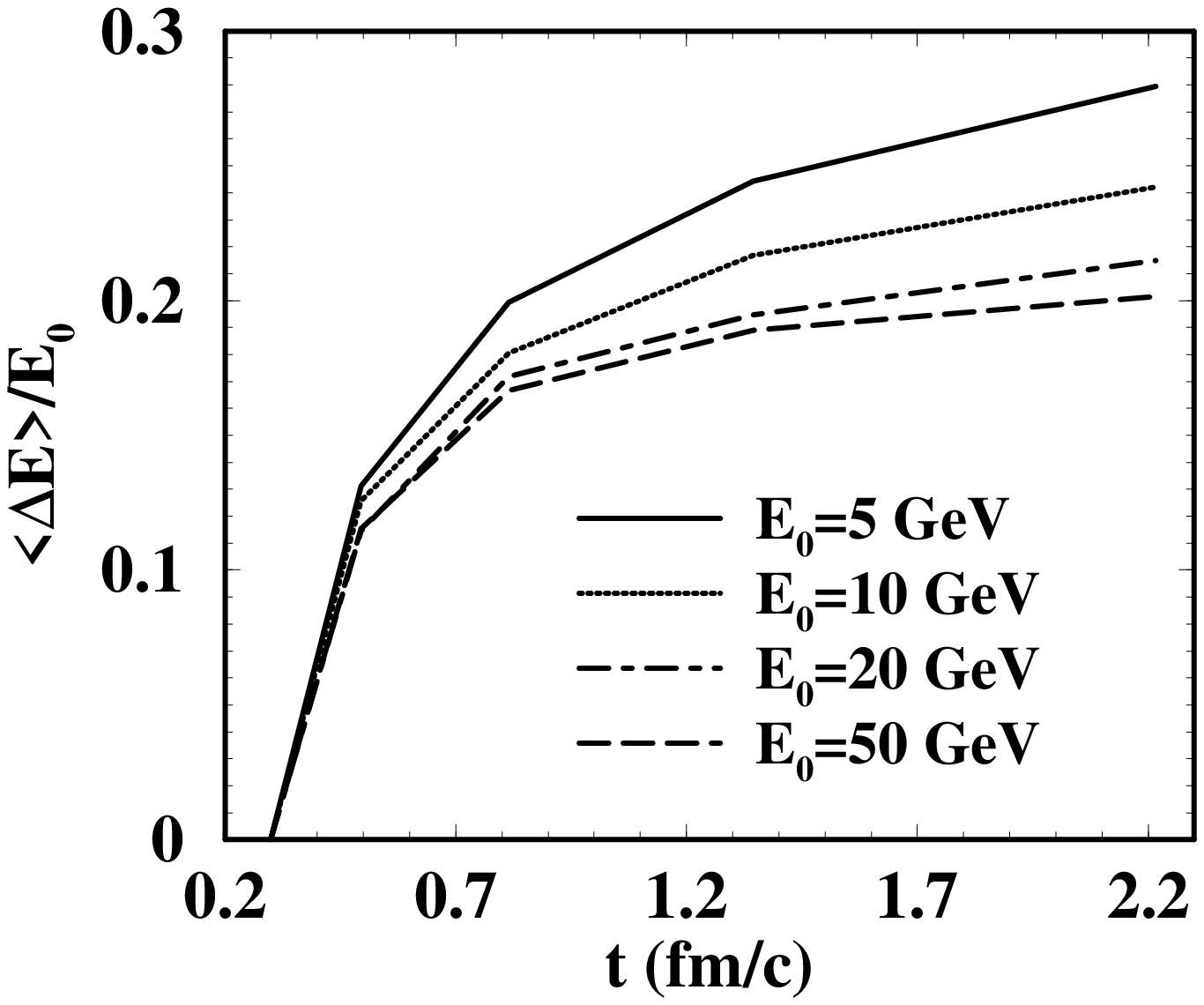, height= 6 cm}
\caption{ Average energy loss of quarks as a function of time for various
initial average energies. 
}
\label{fig1}
\end{figure}
In Fig.~1, we present results for the  energy loss as a
function of time for various initial energies with means indicated in
the legend. It might be noted that at early times fractional energy loss
is independent of the average incoming energy.

\begin{figure}
\epsfig{figure=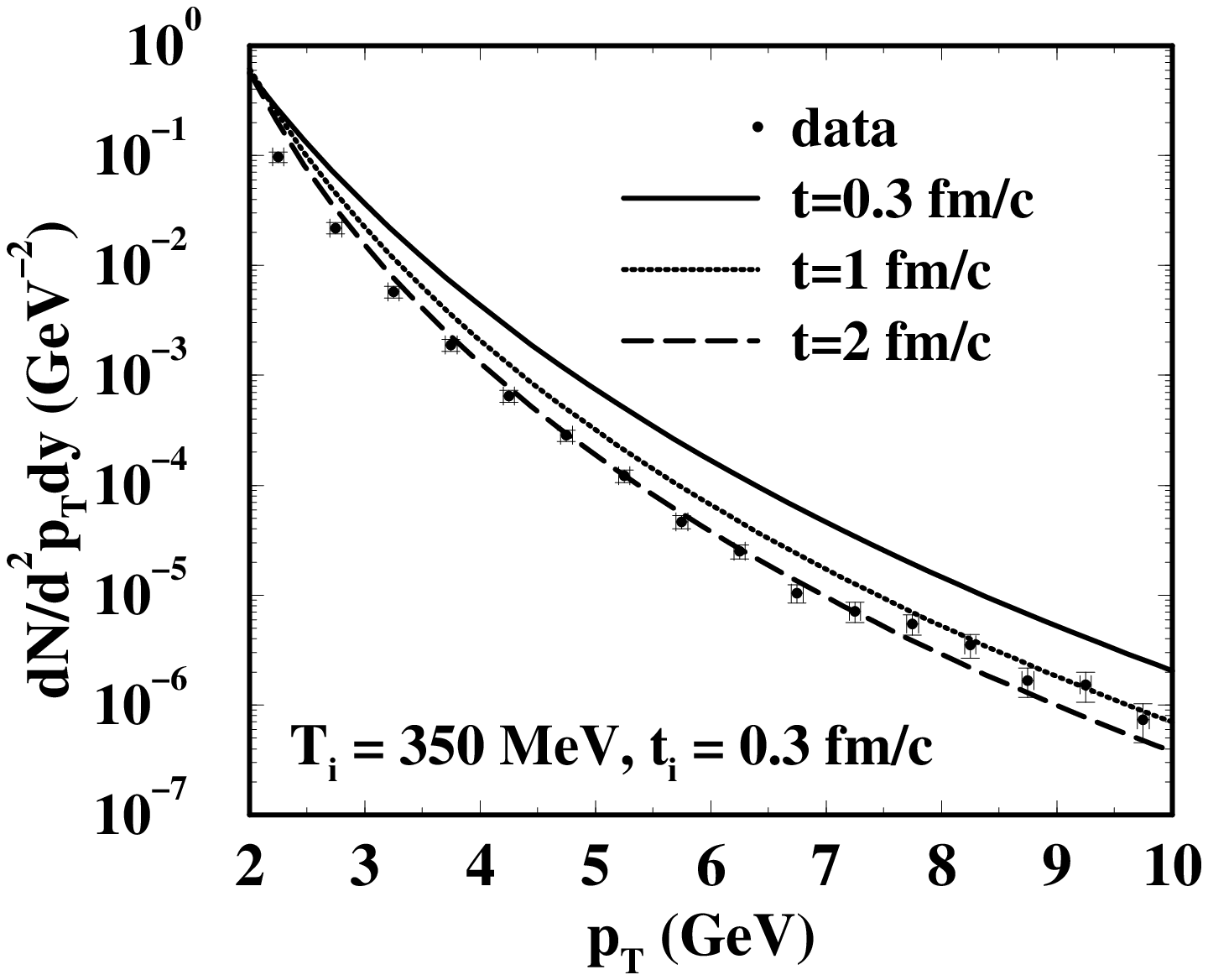, height= 6 cm}
\caption{Time evolution of the transverse momentum  distributions. 
}
\label{fig2}
\end{figure}

Next we investigate time evolution of quark $p_T$ spectra. 
Assuming that the  high $p_T$ partons materialize into
hadrons outside the system and the hadronization process does not
affect the shape of the $p_T$ distribution drastically. Under these
circumstances the pion $p_T$ spectra may be taken to be proportional to
quark $p_T$ spectra for a given time.
Therefore, we take the 
initial $p_T$ distribution of quark
to be proportional to the  pion $p_T$ spectrum  as 
measured in $p$-$p$ collision~\cite{phenixpppi0}:
\bea
f(p_T,p_z,t=t_i)&=&\frac{N_0}{p_T}\frac{dN(y=0)}{d^2p_Tdy}\nn\\
              &=&\frac{\bar{N}_0}{p_T}\frac{1}{(1+\frac{p_T}{p_0})^\nu},
\eea
where, $\nu=9.97$ and $p_0=1.212$ and $N_0 ({\bar N}_0)$
 is the normalization constant. 
The final spectrum is obtained by assuming
the transition temperature, $T_c\sim 190$ MeV at a time 
$t\sim 2$ fm/c. The results are compared with the measured   
$p_T$ spectrum of pions (for $p_T\gtrsim 2$ GeV) at RHIC~\cite{auaupi0} 
in Fig.\ref{fig2}.  The data is well described for $p_T=3-8$ GeV.

In conclusion, in the present work, we estimate quark energy loss
considering stochastic
nature of the interaction by solving FP equation. 
Relevant drag and diffusion coefficients have been calculated in the soft 
collision limit. To highlight the importance of collisional energy loss 
radiative processes is excluded. We have analysed the pion $p_T$ spectrum
measured at RHIC by the PHENIX collaboration~\cite{auaupi0}.  The two body 
scattering is found to give reasonable amount of quenching required to  
explain the data. Recently transverse momentum spectrum
for the $D$ meson has also been measured via single electron $p_T$ distribution
\cite{kharzeev,adler05}.  The present
formalism can be applied to analyse the data reported in ref.~\cite{adler05}. 
Such investigations are currently under progress \cite{alamD}.

\end{multicols}
\end{document}